\documentclass[aps,twocolumn,groupedaddress,showpacs]{revtex4}
\usepackage{graphicx}
\usepackage{latexsym}
\usepackage{amssymb}
\usepackage{amsfonts}
\begin{document}
\bibliographystyle{apsrev}
\title{Nanoengineered magnetic-field-induced superconductivity}
\author{Martin Lange}
\email{martin.lange@fys.kuleuven.ac.be}
\author{Margriet J. Van Bael}
\author{Yvan Bruynseraede}
\author{Victor V. Moshchalkov}
\affiliation{Laboratorium voor Vaste-Stoffysica en Magnetisme, K.
U. Leuven, Celestijnenlaan 200D, 3001 Leuven, Belgium}
\date{\today}
\begin{abstract}
The perpendicular critical fields of a superconducting film have been strongly
enhanced by using a nanoengineered lattice of magnetic dots (dipoles) on top of the
film. Magnetic-field-induced superconductivity is observed in these hybrid
superconductor / ferromagnet systems due to the compensation of the applied field
between the dots by the stray field of the dipole array. By switching between
different magnetic states of the nanoengineered field compensator, the critical
parameters of the superconductor can be effectively controlled.
\end{abstract}

\maketitle

When the applied magnetic field exceeds a certain critical value, superconductivity is
suppressed due to orbital and spin pair breaking effects. This very general property
of superconductors sets strong limits for their practical applications, since, in
addition to applied magnetic fields, the current sent through a superconductor also
generates magnetic fields, which can lead to a loss of zero resistance. Materials that
are not only able to withstand magnetic fields, but in which superconductivity can
even be induced by applying a magnetic field, are very rare and up to now only
(EuSn)Mo$_6$S$_8$ \cite{wolf1982,meul1984}, organic $\lambda$-(BETS)$_{2}$FeCl$_{4}$
materials \cite{uji2001,balicas2001} and HoMo$_6$S$_8$\cite{giroud1987} show this
unusual behavior. The appearance of magnetic-field-induced superconductivity (FIS) in
the former two compounds was interpreted in terms of the Jaccarino-Peter effect
\cite{jaccarino1962}, in which the exchange fields from the paramagnetic ions
compensate an applied magnetic field, so that the destructive action of the field is
neutralized.\\ Here we report that FIS can also be realized in hybrid superconductor /
ferromagnet nanostructured bilayers.
\begin{figure}
 \includegraphics{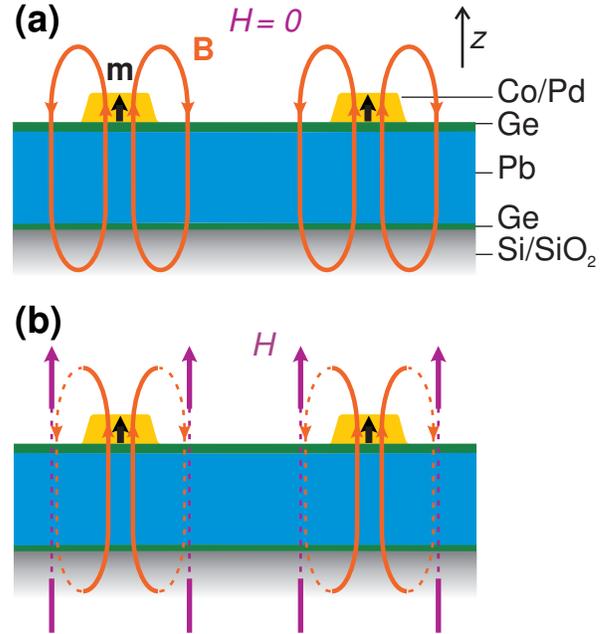}
\caption{Schematic drawing of the investigated hybrid superconductor / ferromagnet
sample. (a) The magnetic stray field $B$ of the dots is comparable with the field of a
magnetic dipole. (b) A magnetic field $H$ applied in the $z$-direction can be
compensated by the dipole stray field between the dots, resulting in the conditions
necessary for the observation of magnetic field-induced superconductivity.}
 \label{fig1}
\end{figure}
The basic idea is quite straightforward (see Fig.~\ref{fig1}): a lattice of magnetic
dots with magnetic moments aligned along the positive $z$-direction is placed on top
of a superconducting film. The magnetic stray field of each dot has a positive
$z$-component of the magnetic induction $B_{z}$ under the dots and a negative one in
the area between the dots. Added to a homogeneous magnetic field $H$, see
Fig.~\ref{fig1}(b), these dipole fields {\em enhance} the $z$-component of the
effective magnetic field $\mu_{0} H_{eff} = \mu_{0} H + B_{z}$ in the small area just
{\em under the dots} and, at the expense of that, {\em reduce $H_{eff}$ everywhere
else} in the Pb film, thus providing the condition necessary for the FIS observation.
This new field compensation effect is not restricted to specific superconductors, so
that FIS could be achieved in any superconducting film with a lattice of magnetic
dots.\\ To implement the idea of the nanoengineered FIS, we have prepared a sample,
which reminds us of other systems used during the last decade for studying flux
pinning by periodic arrays of magnetic dots \cite{geoffrey1993, martin1997,
morgan1998, vanbael1999, vanbael2000}, and by magnetic domains \cite{lange2002}. The
sample consists of a 85~nm superconducting Pb film evaporated on a 1~nm Ge base layer
on an amorphous Si/SiO$_{2}$ substrate, which is held at liquid nitrogen temperature
during deposition. This thin Pb-film behaves as a type-II superconductor. For
protection against oxidation, the Pb is covered by a 10~nm Ge layer that is insulating
at low temperatures and thus prevents the influence of the proximity effect between Pb
and Co/Pd. The Ge/Pb/Ge trilayer is patterned into a transport bridge (width
200~$\mu$m, distance between voltage contacts 630~$\mu$m) using optical lithography
and chemical wet etching. The ferromagnetic dots are made by defining a resist mask on
the transport bridge by electron-beam lithography and subsequent evaporation of a
Pd(3.5~nm)/[Co(0.4~nm)/Pd(1.4~nm)]$_{10}$ multilayer into the resist mask. The resist
is finally removed in a lift-off procedure. The dots are arranged in a regular square
array with period 1.5~$\mu$m. They have a square shape (side length about 0.8~$\mu$m)
with slightly irregular edges.\\ The dots on the superconducting Pb film consist of
Co/Pd multilayers having an easy axis of magnetization perpendicular to the sample
surface \cite{carcia1985}. The hysteresis loop of the dots is measured with $H$
perpendicular to the surface by magneto-optical Kerr effect, revealing a high magnetic
remanance of $M_{r} = 0.8 M_{s}$, where $M_{r}$ and $M_{s}$ are the remanent and
saturation magnetization, respectively, and a large coercive field $\mu_{0} H_{coe} =
150$~mT. This makes it possible to produce quite stable remanent magnetic domain
states in the dots by using different magnetization procedures. These domain states
were investigated by magnetic force microscopy (MFM) in a Digital Instruments
nanoscope III.
\begin{figure}
\includegraphics{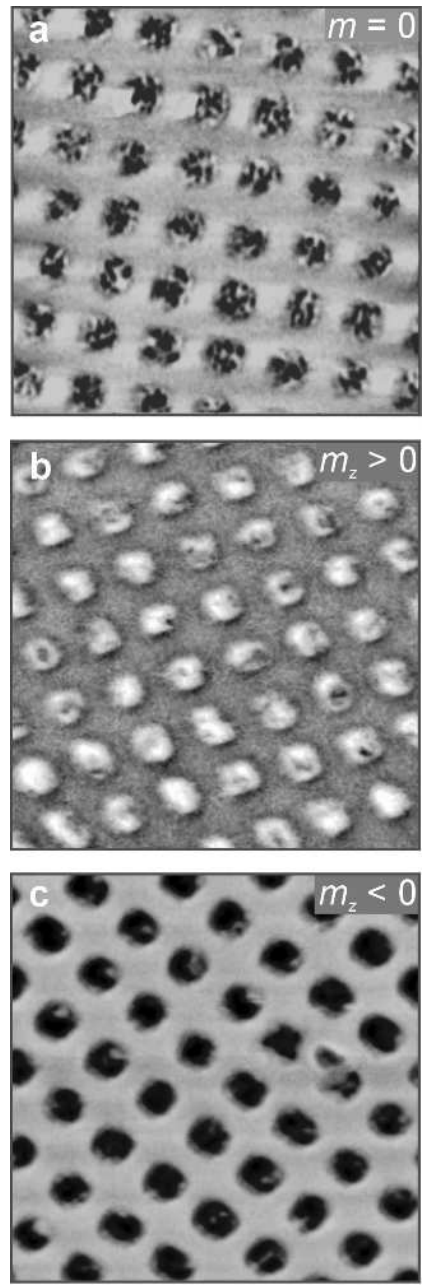}
\caption{MFM images of the hybrid superconductor / ferromagnet structure in $H=0$ and
at room temperature. The images show a $10 \times 10 \mu$m$^{2}$ region of the sample
in the remanent state after (a)~demagnetization, (b)~magnetization in $H = +1$~T,
(c)~magnetization in $H = -1$~T.}
 \label{fig2}
\end{figure}
After demagnetization, the signal from each of the dots consists of dark and bright
spots, as shown in Fig.~\ref{fig2}(a), indicating the presence of several magnetic
domains in the dots, compare Ref.~\cite{hehn1996}. The net magnetic moment $m$ of each
dot in this state is approximately zero ($m=|\mathbf{m}|=0$). The demagnetization is
carried out by oscillating $H$ (perpendicular to the sample surface) around zero with
decreasing amplitude. Saturating the dots in a large positive perpendicular field
aligns all $m$ along the positive $z$-direction ($m_{z}>0$), so that the dots appear
brighter compared with the signal between the dots, see Fig.~\ref{fig2}(b). In
contrast to that, when the dots have been saturated in a large negative field,
resulting in $m_{z}<0$, they give a darker contrast in the MFM image, as shown in
Fig.~\ref{fig2}(c). Simultaneous recording of magnetic and topographic images shows
that the spots visible on dots in (b) and (c) are of topographic origin, and are not
due to a magnetic
signal.\\
\begin{figure*}
\includegraphics{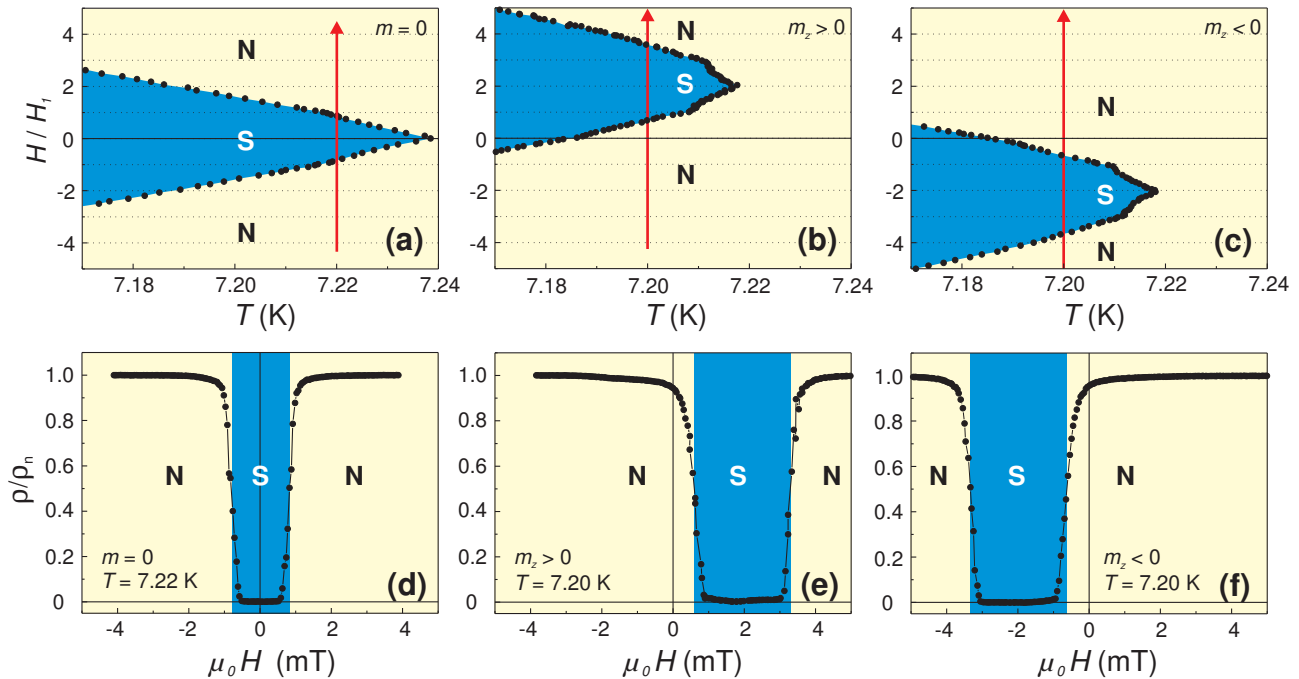}
 \caption{Field - induced superconductivity (FIS) in a Pb film
with an array of magnetic dots. Blue and yellow areas correspond to the
superconducting (S) and the normal state (N), respectively. The $H$-$T$-phase diagrams
are obtained after (a)~demagnetization ($m=0$), (b)~saturation of the dots in a large
positive $H$ ($m_{z}>0$), (c)~saturation in a large negative $H$ ($m_{z}<0$). $\rho
(H)$ is shown for the magnetic states (d)~$m=0$, (e)~$m_{z}>0$, and (f)~$m_{z}<0$,
measured at lines of constant $T$ as indicated by red arrows in the corresponding
phase diagrams.}
 \label{fig3}
\end{figure*}
The magnetic field ($H$)- temperature ($T$) - phase diagrams of the Pb film were
constructed for the three magnetic states of the dots from $\rho(T)$-measurements
carried out in a Quantum Design Physical Properties Measurement System applying a
4-probe ac technique with an ac-current of 10~$\mu$A at a frequency of 19~Hz. $H$ is
applied perpendicular to the sample surface. We defined the critical temperature as
$T_{c} = T(\rho = 50\% \rho_{n})$, with $\rho$ the resistivity and
$\rho_{n}$~=~1.4~$\mu\Omega$~cm the normal state resistivity at 7.3~K. We did not
observe any indication that the small magnetic fields $|H| \ll H_{coe}$ applied during
these measurements altered the domain state of the dots, although minor microscopic
changes cannot be excluded.\\ The $H$-$T$-phase boundary separating the normal (N)
from the superconducting (S) state is clearly altered by changing the magnetic state
of the dot array. A conventional symmetric (with respect to $H$) phase boundary is
obtained when $m=0$, see Fig.~\ref{fig3}(a). Two kinks in the curve can be seen at $H
= \pm H_{1}$, with $H_{1}$ the first matching field $\mu_{0} H_{1} = \phi_{0} / (1.5
\mu{\textrm m})^{2} = 0.92$~mT, at which the applied flux per unit cell of the dot
array is exactly one superconducting flux quantum $\phi_{0} = 2.07$~mT~$\mu$m$^{2}$.
In contrast to that, {\em the $H$-$T$-phase boundary is strongly asymmetric with
respect to $H$ when the dots are magnetized in positive or negative directions}, see
Figs.~\ref{fig3}(b) and \ref{fig3}(c). Moreover, the maximum $T_{c}$ is shifted to $+2
H_{1}$ when $m_{z}>0$ and to $-2 H_{1}$ when $m_{z}<0$. This shift gives rise to FIS
when $m_{z}>0$ and $m_{z}<0$, as is demonstrated in Figs.~\ref{fig3}(e) and
\ref{fig3}(f). For instance, for $m_{z}>0$ and $T = 7.20$~K, the sample is in the
normal state in zero field, but when a positive field between $+0.6$~mT and $+3.3$~mT
is applied, the Pb film becomes superconducting, as shown in Fig.~\ref{fig3}(e).
Similarly, when the magnetic state of the dots is switched to $m_{z}<0$,
superconductivity is induced by applying a negative field between $-3.3$~mT and
$-0.6$~mT, see Fig.~\ref{fig3}(f). Contrary to that, the $\rho(H)$ curve shows the
typical NSN transition of a conventional superconductor for $m=0$.\\
\begin{figure}
\includegraphics{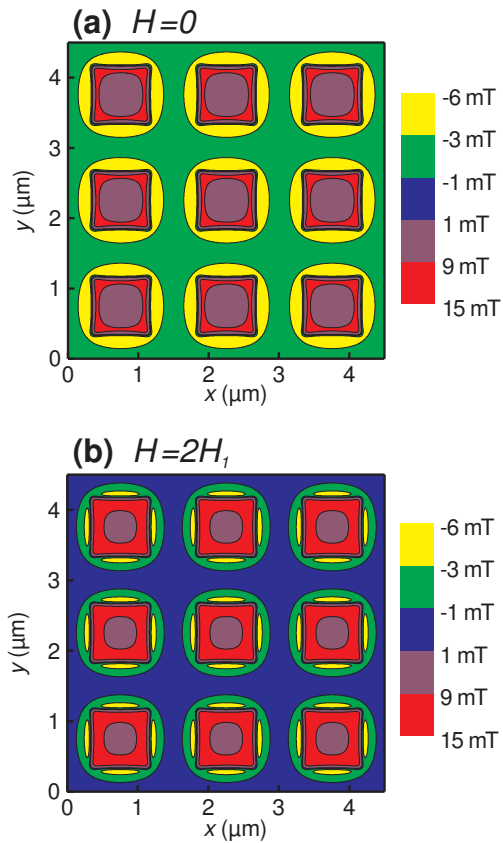}
\caption{Contour plots of the $z$-component of the effective magnetic field $\mu_{0}
H_{eff} = \mu_{0} H + B_{z}$ in the superconductor, calculated using a magnetostatic
model, for (a)~$H=0$, (b) for $H=+2H_{1}$.}
 \label{fig4}
\end{figure}
In the present system, the FIS can be explained by taking into account the local
magnetic induction of the dots $\mathbf{B}$, see the discussion of Fig.~\ref{fig1}. To
support further these arguments, we give in Fig.~\ref{fig4}(a) the distribution of
$B_{z}$, the $z$-component of $B$ for $m_{z}>0$ in the $x$-$y$-plane, calculated by
using a magnetostatic model (see, e.g., Ref.~\cite{jackson}). In zero field
(Fig.~\ref{fig4}(a)) the magnetic dipoles generate stray fields exceeding the upper
critical field of the Pb film when $T > 7.185$~K, and, as a result, the Pb film is in
the normal state. In an applied field of $H=+H_{2}$, the compensation of $B_{z}$ takes
place in the interdot area where the Pb film is now in the superconducting state (see
the blue color in Fig.~\ref{fig4}(b)), thus providing the percolation through
dominantly superconducting areas, and making possible the continuous flow of Cooper
pairs and zero film resistance.\\ An important feature to note here is the appearance
of periodic kinks in the $H$-$T$-phase boundary with a period coinciding with the
first matching field $H_{1}$. These kinks are due to fluxoid quantization effects
\cite{little1962}, confirming that superconductivity indeed nucleates in multiply
connected regions of the film, like in superconducting wire networks \cite{pannetier}
or thin films with periodic arrays of antidots \cite{bezryadin1995,moshchalkov}. The
maximum $T_{c}$ at exactly $H=+2H_{1}$ can therefore be understood in terms of fluxoid
quantization: the flux created by the stray field between the dots can be estimated
from the magnetostatical calculations to be about $-2.1 \phi_{0}$ per unit cell of the
dot array. This makes $H=+2H_{1}$ a favorable field for fulfilling the fluxoid
quantization constraint. Similar arguments can also be applied for the dots in the
$m_{z}<0$ state to explain the shift of the maximum $T_{c}$ to $H=-2H_{1}$. For $m=0$,
$\mathbf{B}$ is strongly reduced due to the domain structure in the dots. This means
that the stray field only weakly influences the Pb film, leading to a phase boundary
without peculiarities except the weak kinks at $H = \pm H_{1}$. To understand all
features in the phase boundaries in more details, one should solve the linearized
Ginzburg-Landau (GL) equations, taking explicitly into account the additional vector
potential created by the magnetic dots. This work has not been done yet, but the GL
analysis of the flux distribution around magnetic dots has been already reported in
Ref.~\cite{milosevic2002}.\\ Moreover, the simple picture of field compensation is not
applicable anymore deeper in the superconducting state. At zero applied field, the
dipole stray field of a magnetic dot generates two vortex-antivortex pairs, with the
vortices located on the dot sites, and the antivortices between the dots. When an
external field is applied, vortices enter the sample and interact with the
vortex-antivortex pairs associated with the dots. This process gives rise to
interesting new effects dealing with vortex-antivortex patterns (see, e.g.,
Ref.~\cite{erdinla}). These novel effects cannot be described in more details here due
to a lack of space, but will be reported elsewhere.\\ The field region in which FIS is
observed can be tuned by changing the period of the dot array or the magnitude of the
stray field. For instance, an increase of the fields emanating from the dots could be
achieved by using magnetic dipoles with larger magnetic moments, shifting the maximum
of $T_{c}$ to an even higher applied field. Good candidates for that are arrays of
nanodots \cite{ross2002} and nanopillars \cite{chou1994}. For instance, dot arrays
with a period of 70~nm have been fabricated \cite{koike2001}, corresponding to $H_{1}
\approx 0.4$~T, which is already a remarkably high field. Besides improving the
critical fields, the dipole array field compensator can also be used to design logical
devices in which superconductivity is controlled by switching between the two
polarities of the magnetized dot array.\\ In conclusion, we have shown that a
nanoengineered lattice of magnetic dipoles can be used to selectively enhance the
critical fields of superconducting films. Magnetic-field-induced superconductivity is
observed due to the compensation of the applied field by the stray field of the
dipoles.\\ The authors are thankful to E. Claessens for help with the measurements,
and to S. Raedts, M. Morelle and K. Temst for their contribution to the sample
preparation. This work was supported by the Belgian IUAP and the Flemish GOA programs,
by the ESF "VORTEX" program, and by the Fund for Scientific Research (F.W.O.) -
Flanders. M.J.V.B. is a Postdoctoral Research Fellow of the F.W.O.-Flanders.

\begin{thebibliography}{24}
\expandafter\ifx\csname natexlab\endcsname\relax\def\natexlab#1{#1}\fi
\expandafter\ifx\csname bibnamefont\endcsname\relax
  \def\bibnamefont#1{#1}\fi
\expandafter\ifx\csname bibfnamefont\endcsname\relax
  \def\bibfnamefont#1{#1}\fi
\expandafter\ifx\csname citenamefont\endcsname\relax
  \def\citenamefont#1{#1}\fi
\expandafter\ifx\csname url\endcsname\relax
  \def\url#1{\texttt{#1}}\fi
\expandafter\ifx\csname urlprefix\endcsname\relax\def\urlprefix{URL }\fi
\providecommand{\bibinfo}[2]{#2} \providecommand{\eprint}[2][]{\url{#2}}

\bibitem[{\citenamefont{Wolf et~al.}(1982)}]{wolf1982}
\bibinfo{author}{\bibfnamefont{S.~A.} \bibnamefont{Wolf}} \bibnamefont{et~al.},
  \bibinfo{journal}{Phys. Rev. B} \textbf{\bibinfo{volume}{25}},
  \bibinfo{pages}{1990} (\bibinfo{year}{1982}).

\bibitem[{\citenamefont{Meul et~al.}(1984)}]{meul1984}
\bibinfo{author}{\bibfnamefont{H.~W.} \bibnamefont{Meul}} \bibnamefont{et~al.},
  \bibinfo{journal}{Phys. Rev. Lett.} \textbf{\bibinfo{volume}{53}},
  \bibinfo{pages}{497} (\bibinfo{year}{1984}).

\bibitem[{\citenamefont{Giroud et~al.}(1987)}]{giroud1987}
\bibinfo{author}{\bibfnamefont{M.}~\bibnamefont{Giroud}} \bibnamefont{et~al.},
  \bibinfo{journal}{J. Low Temp. Phys.} \textbf{\bibinfo{volume}{69}},
  \bibinfo{pages}{419} (\bibinfo{year}{1987}).

\bibitem[{\citenamefont{Uji et~al.}(2001)}]{uji2001}
\bibinfo{author}{\bibfnamefont{S.}~\bibnamefont{Uji}} \bibnamefont{et~al.},
  \bibinfo{journal}{Nature} \textbf{\bibinfo{volume}{410}},
  \bibinfo{pages}{908} (\bibinfo{year}{2001}).

\bibitem[{\citenamefont{Balicas et~al.}(2001)}]{balicas2001}
\bibinfo{author}{\bibfnamefont{L.}~\bibnamefont{Balicas}} \bibnamefont{et~al.},
  \bibinfo{journal}{Phys. Rev. Lett.} \textbf{\bibinfo{volume}{87}},
  \bibinfo{pages}{067002} (\bibinfo{year}{2001}).

\bibitem[{\citenamefont{Jaccarino and Peter}(1962)}]{jaccarino1962}
\bibinfo{author}{\bibfnamefont{V.}~\bibnamefont{Jaccarino}} \bibnamefont{and}
  \bibinfo{author}{\bibfnamefont{M.}~\bibnamefont{Peter}},
  \bibinfo{journal}{Phys. Rev. Lett.} \textbf{\bibinfo{volume}{9}},
  \bibinfo{pages}{290} (\bibinfo{year}{1962}).

\bibitem[{\citenamefont{Geoffroy et~al.}(1993)}]{geoffrey1993}
\bibinfo{author}{\bibfnamefont{O.}~\bibnamefont{Geoffroy}}
  \bibnamefont{et~al.}, \bibinfo{journal}{J. Magn. Magn. Mater.}
  \textbf{\bibinfo{volume}{121}}, \bibinfo{pages}{223} (\bibinfo{year}{1993}).

\bibitem[{\citenamefont{Mart{\'\i}n et~al.}(1997)}]{martin1997}
\bibinfo{author}{\bibfnamefont{J.~I.} \bibnamefont{Mart{\'\i}n}}
  \bibnamefont{et~al.}, \bibinfo{journal}{Phys. Rev. Lett.}
  \textbf{\bibinfo{volume}{79}}, \bibinfo{pages}{1929} (\bibinfo{year}{1997}).

\bibitem[{\citenamefont{Morgan and Ketterson}(1998)}]{morgan1998}
\bibinfo{author}{\bibfnamefont{D.~J.} \bibnamefont{Morgan}} \bibnamefont{and}
  \bibinfo{author}{\bibfnamefont{J.~B.} \bibnamefont{Ketterson}},
  \bibinfo{journal}{Phys. Rev. Lett.} \textbf{\bibinfo{volume}{80}},
  \bibinfo{pages}{3614} (\bibinfo{year}{1998}).

\bibitem[{\citenamefont{Van~Bael et~al.}(1999)}]{vanbael1999}
\bibinfo{author}{\bibfnamefont{M.~J.} \bibnamefont{Van~Bael}}
  \bibnamefont{et~al.}, \bibinfo{journal}{Phys. Rev. B}
  \textbf{\bibinfo{volume}{59}}, \bibinfo{pages}{14674} (\bibinfo{year}{1999}).

\bibitem[{\citenamefont{Van~Bael et~al.}(2000)}]{vanbael2000}
\bibinfo{author}{\bibfnamefont{M.~J.} \bibnamefont{Van~Bael}}
  \bibnamefont{et~al.}, \bibinfo{journal}{Physica C}
  \textbf{\bibinfo{volume}{332}}, \bibinfo{pages}{12} (\bibinfo{year}{2000}).

\bibitem[{\citenamefont{Lange et~al.}(2002)}]{lange2002}
\bibinfo{author}{\bibfnamefont{M.}~\bibnamefont{Lange}} \bibnamefont{et~al.},
  \bibinfo{journal}{Appl. Phys. Lett.} \textbf{\bibinfo{volume}{81}},
  \bibinfo{pages}{322} (\bibinfo{year}{2002}).

\bibitem[{\citenamefont{Carcia et~al.}(1985)\citenamefont{Carcia, Meinhaldt,
  and Suna}}]{carcia1985}
\bibinfo{author}{\bibfnamefont{P.~F.} \bibnamefont{Carcia}},
  \bibinfo{author}{\bibfnamefont{A.~D.} \bibnamefont{Meinhaldt}},
  \bibnamefont{and} \bibinfo{author}{\bibfnamefont{A.}~\bibnamefont{Suna}},
  \bibinfo{journal}{Appl. Phys. Lett.} \textbf{\bibinfo{volume}{47}},
  \bibinfo{pages}{178} (\bibinfo{year}{1985}).

\bibitem[{\citenamefont{Hehn et~al.}(1996)}]{hehn1996}
\bibinfo{author}{\bibfnamefont{M.}~\bibnamefont{Hehn}} \bibnamefont{et~al.},
  \bibinfo{journal}{Science} \textbf{\bibinfo{volume}{272}},
  \bibinfo{pages}{1782} (\bibinfo{year}{1996}).

\bibitem[{\citenamefont{Jackson}(1999)}]{jackson}
\bibinfo{author}{\bibfnamefont{J.~D.} \bibnamefont{Jackson}},
  \emph{\bibinfo{title}{Classical Electrodynamics}}
  (\bibinfo{publisher}{Wiley}, \bibinfo{address}{New York, ed. 3},
  \bibinfo{year}{1999}).

\bibitem[{\citenamefont{Little and Parks}(1962)}]{little1962}
\bibinfo{author}{\bibfnamefont{W.~A.} \bibnamefont{Little}} \bibnamefont{and}
  \bibinfo{author}{\bibfnamefont{R.~D.} \bibnamefont{Parks}},
  \bibinfo{journal}{Phys. Rev. Lett.} \textbf{\bibinfo{volume}{9}},
  \bibinfo{pages}{9} (\bibinfo{year}{1962}).

\bibitem[{\citenamefont{Pannetier et~al.}(1991)}]{pannetier}
\bibinfo{author}{\bibfnamefont{B.}~\bibnamefont{Pannetier}}
  \bibnamefont{et~al.}, in \emph{\bibinfo{booktitle}{Quantum Coherence in
  Mesoscopic Systems}}, edited by
  \bibinfo{editor}{\bibfnamefont{B.}~\bibnamefont{Kramer}}
  (\bibinfo{publisher}{Plenum Press}, \bibinfo{address}{New York},
  \bibinfo{year}{1991}), p. \bibinfo{pages}{457}.

\bibitem[{\citenamefont{Bezryadin and Pannetier}(1995)}]{bezryadin1995}
\bibinfo{author}{\bibfnamefont{A.}~\bibnamefont{Bezryadin}} \bibnamefont{and}
  \bibinfo{author}{\bibfnamefont{B.}~\bibnamefont{Pannetier}},
  \bibinfo{journal}{J. Low. Temp. Phys.} \textbf{\bibinfo{volume}{98}},
  \bibinfo{pages}{251} (\bibinfo{year}{1995}).

\bibitem[{\citenamefont{Moshchalkov et~al.}(2000)}]{moshchalkov}
\bibinfo{author}{\bibfnamefont{V.~V.} \bibnamefont{Moshchalkov}}
  \bibnamefont{et~al.}, in \emph{\bibinfo{booktitle}{Handbook of Nanostructured
  Materials and Nanotechnology}}, edited by
  \bibinfo{editor}{\bibfnamefont{H.~S.} \bibnamefont{Nalwa}}
  (\bibinfo{publisher}{Academic Press}, \bibinfo{address}{San Diego},
  \bibinfo{year}{2000}), vol.~\bibinfo{volume}{3}, chap.~\bibinfo{chapter}{9},
  p. \bibinfo{pages}{451}.

\bibitem[{\citenamefont{Milosevic et~al.}(2002)\citenamefont{Milosevic,
  Yampolskii, and Peeters}}]{milosevic2002}
\bibinfo{author}{\bibfnamefont{M.~V.} \bibnamefont{Milosevic}},
  \bibinfo{author}{\bibfnamefont{S.~V.} \bibnamefont{Yampolskii}},
  \bibnamefont{and} \bibinfo{author}{\bibfnamefont{F.~M.}
  \bibnamefont{Peeters}}, \bibinfo{journal}{Phys. Rev. B}
  \textbf{\bibinfo{volume}{66}}, \bibinfo{pages}{024515}
  (\bibinfo{year}{2002}).

\bibitem[{\citenamefont{Erdin}(2002)}]{erdinla}
\bibinfo{author}{\bibfnamefont{S.}~\bibnamefont{Erdin}} (\bibinfo{year}{2002}),
  \bibinfo{note}{unpublished, cond-mat/0211117}.

\bibitem[{\citenamefont{Ross et~al.}(2002)}]{ross2002}
\bibinfo{author}{\bibfnamefont{C.~A.} \bibnamefont{Ross}} \bibnamefont{et~al.},
  \bibinfo{journal}{J. Appl. Phys.} \textbf{\bibinfo{volume}{91}},
  \bibinfo{pages}{6848} (\bibinfo{year}{2002}).

\bibitem[{\citenamefont{Chou et~al.}(1994)}]{chou1994}
\bibinfo{author}{\bibfnamefont{S.~Y.} \bibnamefont{Chou}} \bibnamefont{et~al.},
  \bibinfo{journal}{J. Appl. Phys.} \textbf{\bibinfo{volume}{76}},
  \bibinfo{pages}{6673} (\bibinfo{year}{1994}).

\bibitem[{\citenamefont{Koike et~al.}(2001)}]{koike2001}
\bibinfo{author}{\bibfnamefont{K.}~\bibnamefont{Koike}} \bibnamefont{et~al.},
  \bibinfo{journal}{Appl. Phys. Lett.} \textbf{\bibinfo{volume}{78}},
  \bibinfo{pages}{784} (\bibinfo{year}{2001}).

\end{thebibliography}

\end{document}